\documentclass[10pt,twocolumn,letterpaper]{article}

\usepackage{cvpr}              
\usepackage{amsmath,amsfonts}
\usepackage{algorithmic}
\usepackage{algorithm}
\usepackage{array}
\usepackage{textcomp}
\usepackage{stfloats}
\usepackage{url}
\usepackage{verbatim}
\usepackage{graphicx}
\usepackage{cite}
\usepackage{xcolor}
\usepackage{multirow}
\usepackage{booktabs}
\usepackage{threeparttable}
\usepackage{caption}
\usepackage{subcaption}
\usepackage{tabularx,booktabs}

\definecolor{cvprblue}{rgb}{0.21,0.49,0.74}

\def\paperID{MTF-4} 
\def\confName{CVPR}
\def\confYear{2024}

\title{Automatic Recognition of Food Ingestion Environment \\ from the AIM-2 Wearable Sensor}

\author{
Yuning Huang,
\and
M A Hassan,
\and
Jiangpeng He,
\and
J. Higgins,
\and
Megan McCrory,
\and
Heather Eicher-Miller,
\and
J. Graham Thomas,
\and
Edward Sazonov,
\and
Fengqing Zhu
}

\begin{document}
\maketitle
\begin{abstract}
Detecting an ingestion environment is an important aspect of monitoring dietary intake. It provides insightful information for dietary assessment. However, it is a challenging problem where human-based reviewing can be tedious, and algorithm-based review suffers from data imbalance and perceptual aliasing problems. To address these issues, we propose a neural network-based method with a two-stage training framework that tactfully combines fine-tuning and transfer learning techniques. Our method is evaluated on a newly collected dataset called ``UA Free Living Study", which uses an egocentric wearable camera, AIM-2 sensor, to simulate food consumption in free-living conditions. The proposed training framework is applied to common neural network backbones, combined with approaches in the general imbalanced classification field. Experimental results on the collected dataset show that our proposed method for automatic ingestion environment recognition successfully addresses the challenging data imbalance problem in the dataset and achieves a promising overall classification accuracy of 96.63\%.
\end{abstract}

\section{Introduction}
\label{sec:intro}
\begin{figure}[t]
\centering
\includegraphics[width=0.4\textwidth]{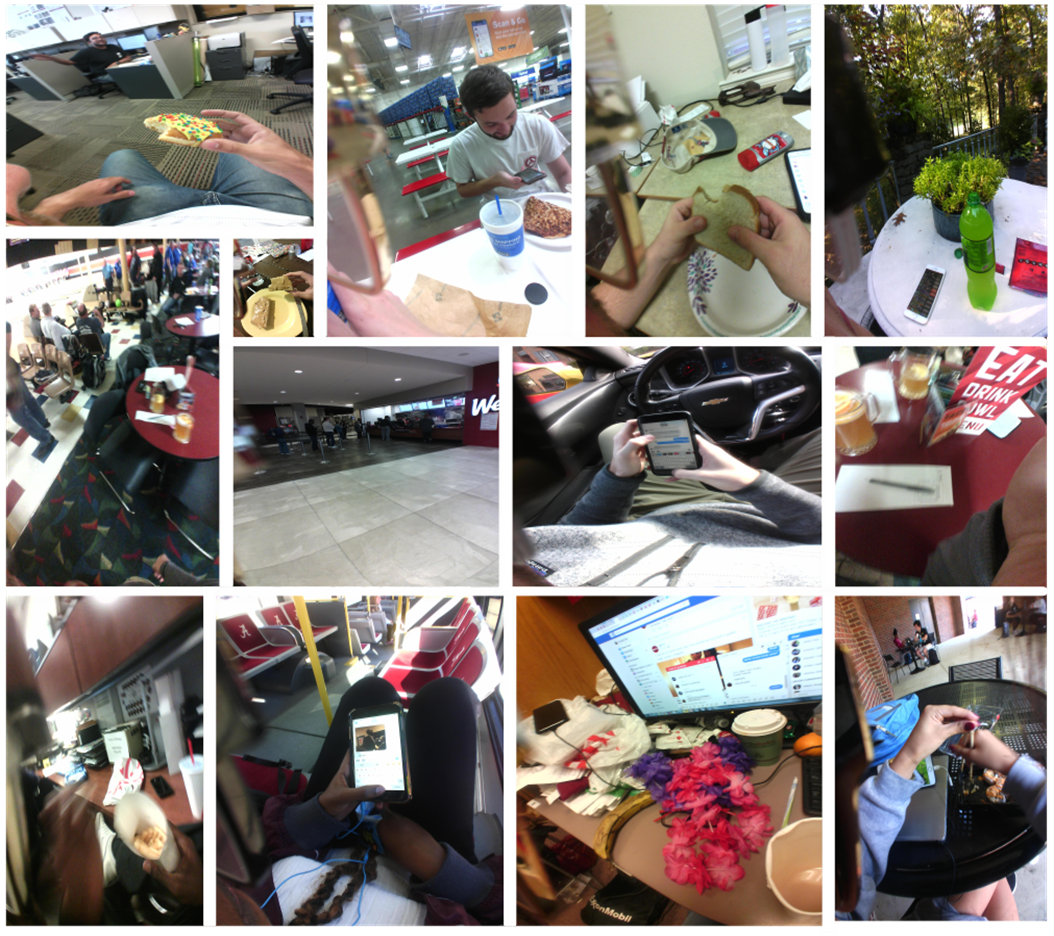}
\caption{Compilation of images showcasing various environments where food may be consumed. The montage is created using free-living data collected from AIM-2. }
\label{fig: figure1}
\end{figure}

Most recent dietary assessment research mainly focuses on monitoring the intake of energy and nutrients in individuals' diets~\cite{sazonov2009toward,farooq2019validation, he2020multitask, shao2023end}. 
However, understanding ingestion behavior, including the impact of the environment and social context, is a relatively new and under-explored aspect of ingestion monitoring. 
The ingestion environment (see Fig.~\ref{fig: figure1}) impacts the dietary behavior of an individual through a range of sensory mechanisms related to food intake~\cite{stroebele2006influence}. For instance,
the environment could influence the quantity of food intake, an individual may consume more foods in one environment compared to another. Likewise, the physical posture during food intake may also vary depending on the ingestion environment. An individual is more likely to sit at a table in a restaurant but may sit at a table, lie on a bed, or sit on a sofa while consuming food at home~\cite{breit2023spectrum}.

The environment can play a vital role in food acceptance. Meiselman et al.~\cite{meiselman2000demonstrations} conduct a study demonstrating the impact of the ingestion environment on food acceptance. The participants were asked to rate their acceptance of the meal by completing a hedonic rating scale. Results show that the meal ratings varied with the environment, \textit{e.g.}, meals are most liked in the restaurant and least liked in the cafeteria. A more recent study~\cite{garcia2015influences} examined how location and table settings affect people's willingness to eat food revealed that participants were substantially more likely to consume food presented on a gourmet table (which is typical in restaurants) than on a home-style table and plastic tray. 

The ingestion environment can also influence the type of food consumed. Bauer et al.~\cite{bauer2022healthy} have shown that food choices made out-of-home are often less consistent with people's dietary plans, and eating out-of-home may be one of the main reasons for the failure of dietary goals. Claessens et al.~\cite{claessens2023personal} also indicate that people's eating choices in restaurants are typically unhealthier and less sustainable than at home; their survey has shown that healthiness is the most important consideration for choosing home meals while being the third most important factor for choosing restaurant meals. Another recent study~\cite{breit2023spectrum} investigated the distribution of eating locations for breakfast, lunch, dinner, and snacks and showed that the study participants tend to have more snacks in vehicles.

Compared to other aspects of measurement such as eating activity detection and dietary composition/energy estimation, ingestion environment recognition is still an underexplored area. A relevant study by Gemming et al.~\cite{gemming2015use} used a wearable camera to capture and categorize environmental and social contexts to understand food intake behavior. The authors used the SenseCam wearable camera~\cite{van2008faking} to capture eating episodes by monitoring activity throughout the day. The images of the eating episode were manually annotated by two researchers, marking the eating location, external environment, physical position, social interaction, and viewing of media screens. The study reported that the duration of food intake varied with different food locations, such as home, workplace, and restaurant whereas the longest eating episodes tended to occur in restaurants and the shortest in the workplace.

An analysis of many previous studies is conducted by manually reviewing self-report questionnaires and images, however, this may not be an optimal solution. Self-reported data are subject to misreporting. A study from Thea and Mortel~\cite{van2008faking} reported that participants tend to present a favorable image of themselves when completing questionnaires that elicit an evaluative response. In addition, manual review of dietary images~\cite{shao2021_ibdasystem, shao2021_nutri_database} is time-consuming and subject to human error. Furthermore, egocentric wearable cameras such as eButton~\cite{sun2014ebutton} and SensCam operate throughout the day, capturing 4,000 to 6,000 images, while only less than 20\% of the images are relevant to eating events. For all of these egocentric cameras, an automatic method for ingestion environment recognition is needed to facilitate further studies on the influence of the environment on dietary activities.

To successfully design a framework for the eating environment recognition, we need to first define the major challenges within the task. In comparison to the standard environment recognition problem, eating environment recognition has majorly suffered from limited publicly available datasets, perceptual aliasing, and most importantly, severe data imbalance.
Although there are more public datasets for food-related tasks, they seldom provide the environment ground truth labels while food environment-related datasets are hard to get published due to privacy concerns with the egocentric view. In addition, images of different environments may look similar~\cite{dubourg2016sensecam}, such as the dining room of a house and the table setting of a restaurant. Also, the eating scene's distribution may be imbalanced~\cite{breit2023spectrum}, with most consumption taking place at home and less consumption at other locations. Furthermore, the first two issues could exacerbate the impact of data imbalance because the limited dataset size and significant inter-class similarity impose a greater burden on the training of a classifier.

A recent study of automatic eating environment classification~\cite{martinez2019hierarchical} has proposed a VGG network-based hierarchical method to automatically classify 15 food-related scenes with an overall accuracy of 56\%.  However, their method still requires a manual selection of eating episodes from recorded day photo streams and fails to address the challenges mentioned above.

In this paper, we propose to address the recognition of the ingestion environment by automatically detecting eating episodes through the use of a food intake sensor, AIM-2~\cite{doulah2020automatic}, and the associated neural network-based classification framework that specifically targets the aforementioned challenges. 

Our contributions are summarized as follows:
\begin{enumerate}
\item{We identify major challenges in ingestion environment recognition and propose a simple, general, and effective deep-learning-based framework for addressing them.}
\item{We collect and analyze a dataset, \textit{UA Free Living Study}, comprised of unrestricted ad-libitum food consumption for automatic recognition of ingestion environment using an egocentric wearable sensor.}
\item{We conduct various experiments on the collected dataset to verify the performance of the proposed method and show the advantage over generic techniques for imbalanced classification.}
\end{enumerate}

\begin{figure}[t]
\centering
\includegraphics[width=0.4\textwidth]{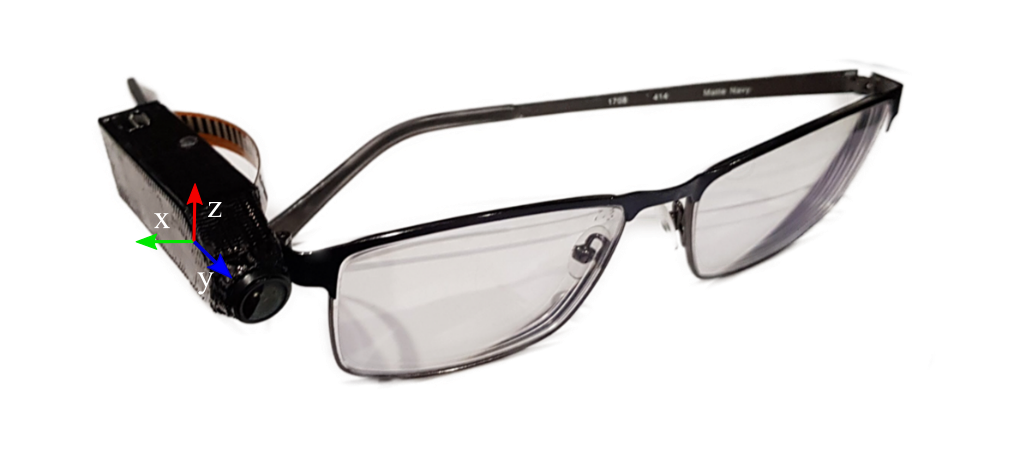}
\flushright
\caption{AIM-2, an egocentric wearable camera that monitors ingestion behavior.}
\label{fig: AIM device}
\end{figure}

\section{Related works}
\label{sec:related works}

\subsection{Environment Recognition}
Multiple computer vision/early deep learning algorithms related to environment recognition were reviewed in~\cite{lowry2015visual}. Before the era of deep learning, researchers in traditional computer vision and robotics attempted to address the issue of perceptual aliasing by using local and global feature descriptors. 
The local descriptors operate as a two-stage process of feature extraction and recognition.  Feature descriptors such as SIFT~\cite{lowe1999object} (scale-invariant feature transform) and SURF~\cite{bay2006surf} (speeded-up robust features) are used to extract features and detect objects in an image. 
Similarly, global feature descriptor-based methods also operate as a two-stage process. Feature descriptors such as color histograms, descriptor-based PCA~\cite{ke2004pca} (principal component analysis), and histogram of oriented gradients~\cite{deniz2011face} methods are used to process the entire image and capture edges, corners, and color patches. The features are processed using machine learning methods such as SVM~\cite{cortes1995support} (support vector machine).
However, with the progress of network design and training, neural network-based environment recognition has achieved better accuracy than traditional methods since it is more powerful at effectively extracting classification-related features. Multiple network architectures such as VGG16~\cite{simonyan2015vgg}, ResNet~\cite{he2016resnet}, ConvNeXt~\cite{liu2022convnext} and Vision Transfomer~\cite{dosovitskiy2020image,liu2021swin} may be used to address environment recognition and classification problem. Among these architectures, the transformer-based methods have achieved the best overall performance. In this work, we select three representative network architectures as the backbone of the proposed framework, which are ResNet, ConvNeXt, and Swin transformer.

\subsection{Imbalanced Classification}
Imbalanced data is a challenging issue for classification where the number of samples belonging to different classes is largely different~\cite{he2022long,cao2019learning,wang2021longtailed}. This problem, if not carefully addressed, often leads to unsatisfactory prediction accuracy on test data, especially for minority classes (classes that contain fewer samples). There are many general approaches proposed for training deep networks on an imbalanced dataset. Class re-balancing is a major paradigm in imbalanced learning and there are two widely-used effective categories of methods that belong to this paradigm, one is resampling~\cite{chawla2002smote,liu2008exploratory,zhang2021learning, he_2023icip, he2023long} and another is weighted loss function~\cite{lin2017focal,cui2019class,tan2020equalization, mao2020visual}. The resampling strategy is basically to re-assign the probability of the sample from each class to be presented in the training mini-batch. Samples from the minority class will appear with a higher probability while samples from the majority class will appear with a lower probability. The weighted loss function is used to adjust the training loss values for different classes by assigning them different weights corresponding to the number of samples in the class. The weighted loss has larger gradients from minority class samples and thus encourages the model to be more adaptive for learning features from minority classes. There are also other efforts for solving this challenging problem such as information augmentation~\cite{kim2020m2m} and improvement on network module design~\cite{huang2016learning}. In this work, we adopt two most widely used approaches, random resampling and weighted cross-entropy loss. We conduct experiments to verify our proposed framework has achieved a better performance boost while also maintaining a positive interaction with both approaches.
\section{Dataset}
\label{sec:datasetcollection}
In this section, we introduce how we collect and construct the dataset we need to perform automatic recognition of eating events, such as meals and snacks.
\subsection{Sensor System}
The sensor system used for the method development is the Automatic Ingestion Monitor, version 2 (AIM-2)~\cite{doulah2020automatic}, a second-generation egocentric wearable (Fig.~\ref{fig: AIM device}) for monitoring dietary intake and eating behavior. 
\subsection{Data Collection}
We collected experimental data from thirty volunteer participants (65\% males and 35\% females, aged 18 to 39 years old). The University of Alabama institutional review board approved the study, and participants were compensated for their participation. The subjects represented four races, non-Hispanic, African American, Asian, and Hispanic. The experiment was conducted in two parts: a controlled laboratory experiment and a free-living experiment. 

For this study, we used the data from the free-living experiment. The participants were asked to wear the AIM-2 sensor for the entire day, follow their normal daily activities, and have at least a single meal at a place of their choice. The participants wore the device for 8.5 to 15.75 hours. The participants were not limited to any social/personal interaction, activities (except for those they considered private or water-based activities), consumption of particular food types, or how the food type was consumed. 

The participants were asked to self-report all eating events (both solids and liquids) using the ASA24~\cite{kupis2019assessing} (Automated Self-Administered 24) in food diary mode after completing the day of AIM-2 monitoring during free-living.

In total, the participants consumed 89 meals in four different environments: vehicle, home, restaurant, and workplace. We corrected the falsely reported self-assessment data by using the self-reporting data correction approach described by Giacchi et al.~\cite{giacchi1998correction}. We performed the expert review for the entire dataset as our population was significantly smaller compared to the population reported in~\cite{giacchi1998correction}. During the expert review, we reviewed each image sequence including images before, after, and during the eating episode to determine the actual ingestion environment. After this, we made corrections to the self-reported ingestion environment, see more details of the construction of the dataset in supplementary materials.

\begin{table}[]
\caption{Sample distribution in the dataset}
\resizebox{0.475\textwidth}{!}{%
\begin{tabular}{l|c|c|c|c}
\hline
                        & Vehicle & Home   & Restaurant & Workplace \\ \hline
Percentage at seq-level & 2.2\%   & 71.9\% & 11.3\%     & 14.6\%    \\ \hline
Percentage at img-level & 2.0\%   & 67.0\% & 13.6\%     & 17.4\%    \\ \hline
\end{tabular}%
}
\label{table:sample distribution}
\end{table}

\subsection{Statistics and Challenges}
\label{sec: challenge and limitation}
After labeling the eating environment, we constructed the dataset for training and testing the proposed method.
However, as we discussed in Section~\ref{sec:intro}, the ingestion environment recognition task contains several inherent issues that require specific considerations in designing methods for it.

First, the number of samples in each class is very imbalanced (see Table~\ref{table:sample distribution}) with the ratio between the major class and minor class being 35. This severe data-imbalanced problem is an inherent issue as most people consume more meals at home and consume fewer meals in other places. The imbalance issue of the dataset introduces unfair favors of the classifier towards the majority class and ignorance toward the minority class.

Second, the total number of sequences and images collected in our dataset is limited, with only 89 sequences and 5,351 images. The size of the dataset makes the imbalance issue more difficult to handle since normal methods may introduce overfitting in training toward the minority class.

In addition, the dataset is egocentric, which is different than general classification datasets (e.g. ImageNet~\cite{deng2009imagenet} and Place365~\cite{zhou2017places}). The egocentric sequence may bear great inter-class similarity due to their viewpoint. For example, when the participant is looking directly at the food, it may be hard to distinguish between ``Home" and ``Restaurant" since the scene captured may only contain a table with food on it with limited background information.

The aforementioned problems make our dataset even more challenging compared to the normal imbalanced datasets. In the experiments, we show that chosen general techniques for addressing data imbalanced classification may not directly work well for our dataset.

In Section~\ref{sec:recognition}, we introduce an automatic scene recognition method with a two-stage training framework that addresses the challenging issue in the proposed dataset.

\begin{figure*}[t]
\centering
\includegraphics[width=0.95\textwidth]{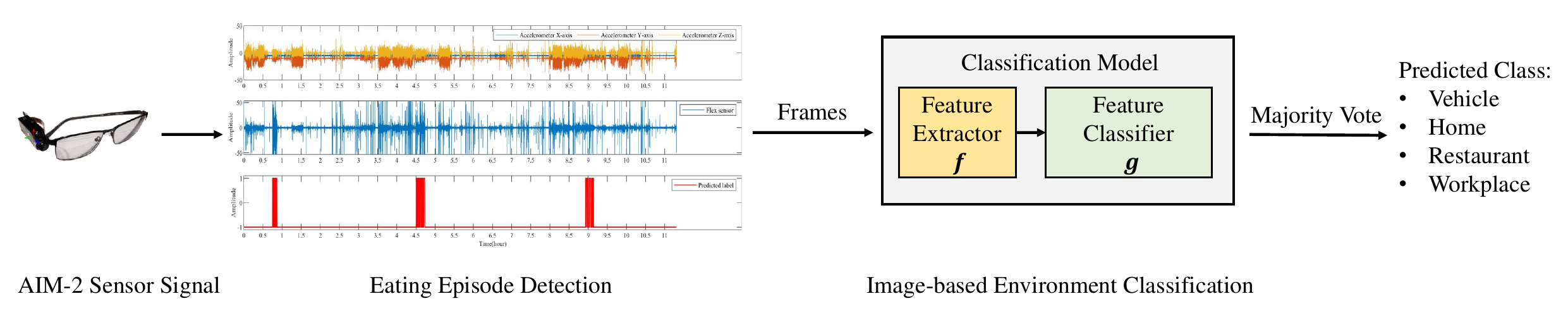}
\caption{Description of the proposed method for automatic ingestion environment recognition.}
\label{fig: overallarch}
\end{figure*}
\section{Method}
\label{sec:recognition}

In this paper, we choose to perform neural network-based scene recognition. We select ResNet~\cite{he2016resnet}, ConvNeXt~\cite{liu2022convnet} and Swin transformer~\cite{liu2021swin} as our representative network backbones. Note that our method is not restricted to the special design of any general classification architectures and thus can be easily adapted to other network backbones.

\subsection{Overall Design}
The overall architecture of the proposed automatic scene recognition method is shown in Fig.~\ref{fig: overallarch}. We use sensor-captured signals to determine the start time and end time of eating episodes~\cite{farooq2018accelerometer,doulah2018importance} (more details in supplementary materials). Note that the network used in the figure is already trained and finetuned with the proposed framework.

As mentioned in Section~\ref{sec: challenge and limitation}, the collected dataset only contains a limited number of samples and has severe data imbalance issues. To address the challenges introduced by the dataset, we propose a two-stage drop-then-maintain training framework that tactfully adopts the techniques from the finetuning and transfer-learning field to achieve balanced training with more training samples.

We propose to drop the feature classifier (see definition in the next sub-section) in the first stage while finetuning the model on Places365 database~\cite{zhou2017places} and maintain the feature classifier while finetuning the model on our dataset, which we referred to as the drop-then-maintain strategy. To enable the second stage of training, we need to additionally process the Place365 database using a semantic-based class filtering and merging.

It is worth noting that the drop strategy is well established and used in the transfer-learning field on the classification task while the maintain strategy is commonly used in other tasks like image restoration when finetuning is useful. However, our proposed drop-then-maintain strategy is less explored, and to the best of our knowledge, the first one to be proposed and applied to the environment recognition field which successfully addresses the inherent challenges mentioned before.

The strategy is designed based on our previous analysis of challenges. In addition, experiments verified its effectiveness against the simple drop strategy or maintain strategy and showed that this tactful combination of the two is helpful and necessary.

\subsection{Method Formulation}
\label{sec: model formulation}
For any general classification network, we can partition them into two major functional parts: feature extractor and feature classifier, where the classifier is the last fully connected layer in the neural network that utilizes highly condensed information extracted by the feature extractor (all previous layers) to perform the classification. The feature extractor can be viewed as a function that maps from image space to feature space, and the feature classifier can be viewed as a function that maps from feature space to label space, which can be formulated as follows: 
\begin{equation}
    f: \mathbb{R}^{H \times W \times 3} \mapsto \mathbb{R}^{C} 
\end{equation}
\begin{equation}
    g: \mathbb{R}^{C} \mapsto \mathbb{R}^{N}
\end{equation}
where $f$ denotes the feature extractor, $g$ denotes the feature classifier that predicts class probabilities, $H$ and $W$ are the height and width of the input image, $C$ is the feature dimension, and $N$ is the number of classes.
Using $f$ and $g$, the classification network can be abstracted as a composed function that maps from image space to label space:
\begin{equation}
    g \circ f: \mathbb{R}^{H \times W \times 3} \mapsto \mathbb{R}^{N}
\end{equation}

In normal transfer learning for classification, the feature classifier $g$ is often dropped and replaced by a new classifier that has the same output dimension as the number of target classes while only the feature extractor is transferred. This is a common practice in the field and has been used in~\cite{zhou2017places} and~\cite{martinez2019hierarchical}.

However, we argue that this is sub-optimal for food environment classification since training new feature classifiers on these datasets suffers a strong bias towards the majority class (\textit{i.e.}, ``Home''). To train a powerful feature extractor and a robust feature classifier, we believe our proposed drop-then-maintain strategy is necessary.

%

In the next sub-section, we will go into the details of our proposed two-stage drop-then-maintain training framework that utilizes semantic-based class filtering and merging.

\begin{figure*}[t]
\centering
\includegraphics[width=0.9\textwidth]{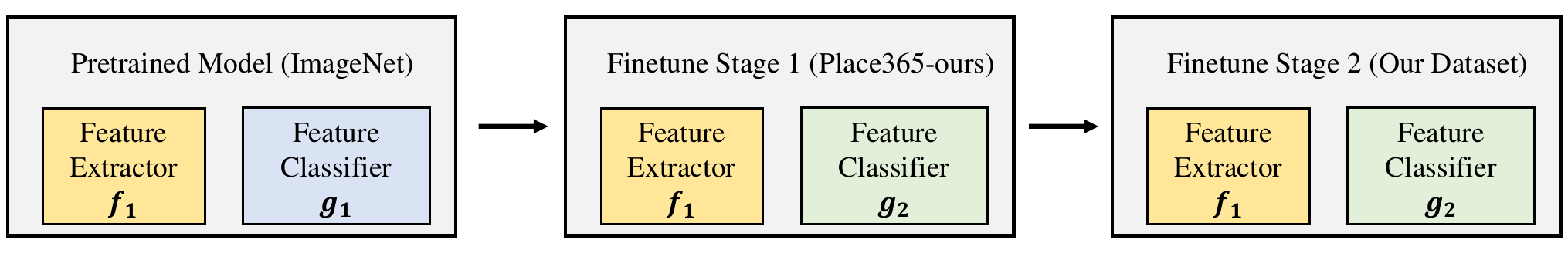}
\caption{Two-stage drop-then-maintain training framework. After the ImageNet pretraining, the feature classifier $g_1$ is dropped and replaced by $g_2$ (since the number of classes has changed from 1,000 to 4), and $g_2$ can remain from stage 1 to stage 2 because of the semantic-based class filtering and merging we performed to Place365 dataset.}
\label{fig: twostagefinetune}
\end{figure*}

\begin{figure}[t]
\centering
\includegraphics[width=0.45\textwidth]{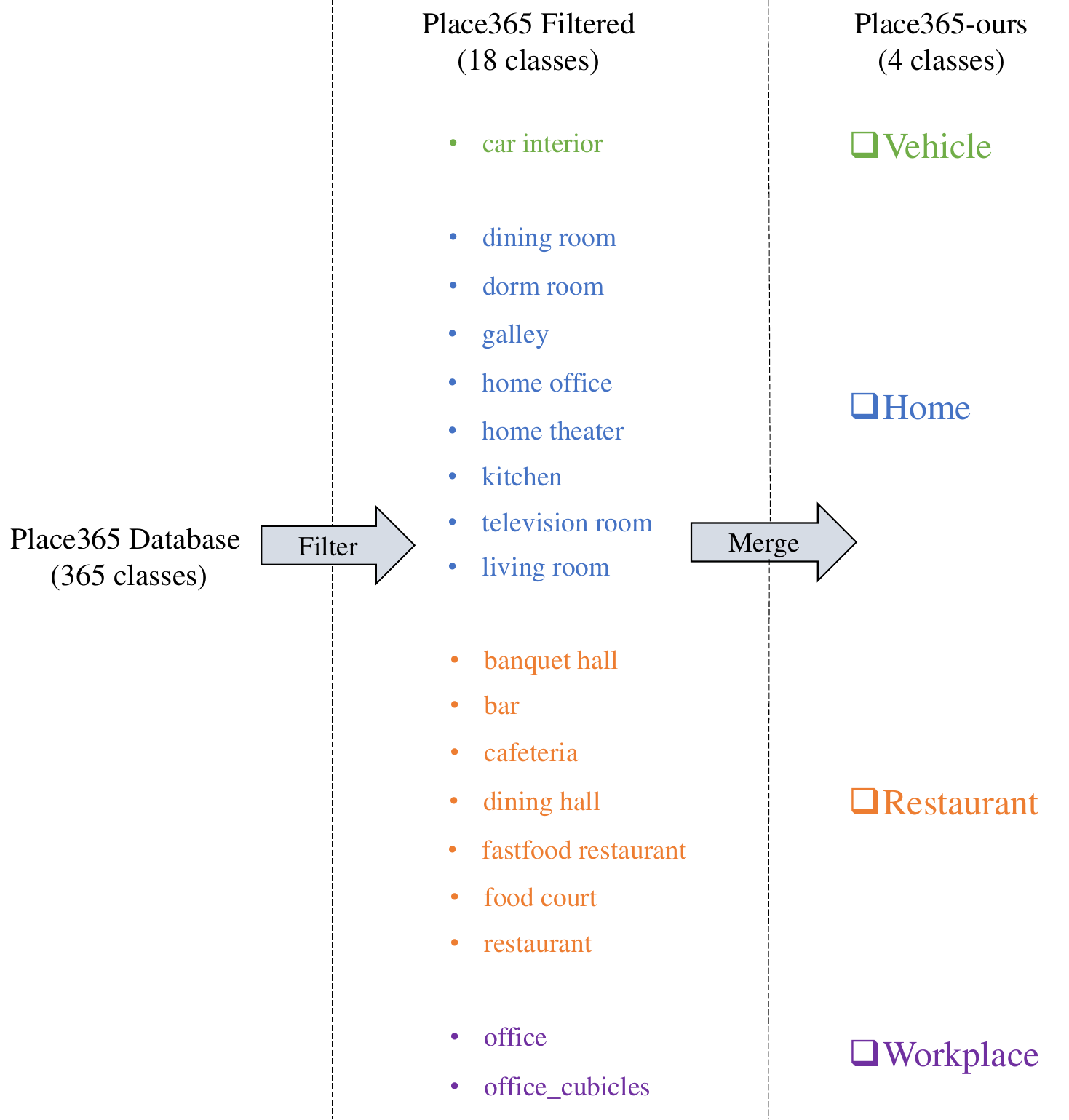}
\caption{Semantic-based dataset filtering and merging.}
\label{fig: semantic heriarchy}
\end{figure}

\subsection{Training Scheme}
To address the data imbalance problem, a straightforward but effective approach is to increase the number of training data~\cite{japkowicz2002class}. We adopt a two-stage training strategy to maximize the utilization of the public dataset to increase the numbers of data in the whole training process, see Fig.~\ref{fig: twostagefinetune}.

\textbf{Dataset selection:} We identify two public datasets that may benefit the performance of the model, one is ImageNet~\cite{li2009imagenet}, and another is Place365~\cite{zhou2017places}.

ImageNet contains 1,000 classes that are different from the four classes we are interested in. Due to its large volume, we consider using the pre-trained model on it since it helps to improve the ability of the feature extractor $f$. It has been shown that pre-training on ImageNet can generally fasten the converging speed of the classification model as well as improve its performance~\cite{huh2016makes}. 

Besides, it is also worthwhile to utilize the Place365 dataset, which is targeted for general scene recognition tasks. Since Place365 contains relevant classes to our target classes, we expect it helpful for improving the performance of feature classifier $g$.
As mentioned before, we want to train $g$ on Place365 and finetune it on our dataset without dropping and retraining a new classifier.
However, the maintaining strategy requires the class match between two datasets and 
the Place365 dataset has many more classes than our dataset. This is where semantic-based class filtering and merging come into play, which ensures both a semantic and dimensional match between the processed Place365 dataset and our collected dataset.

\textbf{Semantic-based class filtering and merging:}
To fulfill the semantic match and keep the output dimension consistent for the feature classifier $g$, we propose a semantic-based class filtering and merging strategy to pre-process the Place365 dataset. First, we identify the relevant classes in the dataset and then discard all the irrelevant classes. After dataset filtering, we merge the remaining classes into four groups in a semantic manner as described in Fig.~\ref{fig: semantic heriarchy}. After filtering and merging, the Place365 dataset has the same number and semantic meaning of classes as our dataset, which we refer to as \textit{Place365-ours}. The Place365-ours is not perfectly balanced but significantly better than our dataset since the ratio between major class and minor class drops from 35 to 8. Besides, the semantic-matched training samples can further increase the intra-class diversity to improve the generalizability of the model.

\textbf{Two-stage drop-then-maintain training:}
To utilize the Imagenet and Place365-ours datasets to improve both the feature extractor $f$ and feature classifier $g$. We first train the model on Place365-ours with an ImageNet pre-trained feature extractor $f_1$ along with a randomly initialized feature classifier $g_2$, note that the ImageNet pre-trained feature classifier $g_1$ is dropped here.
Then, we train the model on our dataset with both the pre-trained feature extractor and feature classifier obtained from the first stage training (see Fig.~\ref{fig: twostagefinetune}). Note that $f_1$ is transferred through both stages while $g_1$ is dropped out at the first stage and $g_2$ is maintained at the second stage. This setting maximizes the utilization of ImageNet and Place365-ours to train a powerful and representative feature extractor and a less biased robust feature classifier.

In both stages of training, we do not freeze the weights of any layers so that the update of the gradient is passed through all layers. We want to finetune both the feature extractor and feature classifier to address the potential domain shift issue between Imagenet, Place365-ours, and our dataset.

\subsection{Approaches for Imbalanced Classification}
We additionally include two approaches for imbalanced classification in our training stage and compare the performance boost in the experiment section.

\textbf{Weighted loss function:} 
We utilized the weighted cross-entropy loss in our model, where the weights are based on the inverse of the sample sizes of each class.

\textbf{Resampling strategy:} We apply the resampling strategy using both over-sampling and under-sampling to ensure the four classes have the same probability of appearing in one training mini-batch.

\begin{table*}[]
\caption{Main Result: Overall Performance}
\centering
\begin{tabular}{c|c|c|c|c|c|c}
\hline
Architectures                    & Model Paras.            & Method        & Seq-level Acc & Macro Precision & Macro Recall & Macro F1 \\ \hline
\multirow{6}{*}{ResNet}          & \multirow{6}{*}{42.51M} & Baseline 0 (DR)     & 88.76         & 67.64           & 59.42        & 62.67    \\ \cline{3-7} 
                                 &                         & Baseline 1 (WL) & 88.76         & 60.90           & 64.01        & 62.04    \\ \cline{3-7} 
                                 &                         & Baseline 2 (RS) & 86.52         & 61.15           & 55.95        & 57.45    \\ \cline{3-7} 
                                 &                         & Ours          & 89.89         & 90.06           & 72.30        & 78.00    \\ \cline{3-7} 
                                 &                         & Ours (WL)       & 91.01         & 90.99           & 74.80        & 80.20    \\ \cline{3-7} 
                                 &                         & Ours (RS)       & \textbf{92.13}         & \textbf{91.58}           & \textbf{76.72}        & \textbf{81.37}    \\ \hline
\multirow{6}{*}{ConvNeXt}        & \multirow{6}{*}{87.58M} & Baseline 0 (DR)     & 89.89         & 64.59           & 63.45        & 63.95    \\ \cline{3-7} 
                                 &                         & Baseline 1 (WL) & 87.64         & 61.65           & 60.56        & 60.76    \\ \cline{3-7} 
                                 &                         & Baseline 2 (RS) & 88.76         & 62.34           & 63.06        & 62.69    \\ \cline{3-7} 
                                 &                         & Ours          & 92.13         & 93.69           & 73.08        & 80.17    \\ \cline{3-7} 
                                 &                         & Ours (WL)       & \textbf{94.38}         & \textbf{94.28}           & \textbf{81.72}        & \textbf{85.82}    \\ \cline{3-7} 
                                 &                         & Ours (RS)      & \textbf{94.38}         & 92.14           & \textbf{81.72}        & 84.54    \\ \hline
\multirow{6}{*}{Swin transformer} & \multirow{6}{*}{86.75M} & Baseline 0 (DR)     & 91.01         & 65.35           & 65.37        & 65.27    \\ \cline{3-7} 
                                 &                         & Baseline 1 (WL) & 91.01         & 64.40           & 65.37        & 64.73    \\ \cline{3-7} 
                                 &                         & Baseline 2 (RS) & 91.01         & 68.66           & 63.84        & 65.91    \\ \cline{3-7} 
                                 &                         & Ours          & 94.38         & 95.29           & 80.19        & 85.34    \\ \cline{3-7} 
                                 &                         & Ours (WL)       & \textbf{96.63}         & \textbf{95.31}           & \textbf{84.61}        & \textbf{87.57}    \\ \cline{3-7} 
                                 &                         & Ours (RS)       & \textbf{96.63}        & \textbf{95.31}           & \textbf{84.61}        & \textbf{87.57}    \\ \hline
\end{tabular}
\label{table:main result 1}
\end{table*}
\section{Experiment and Analysis}
\label{sec:experiment}
In this section, we report the results of our proposed method and compare it to three Baseline methods:

\textbf{Baseline 0 (direct, or DR):} Use ImageNet pre-trained model, directly finetune on our dataset.

\textbf{Baseline 1 (weighted loss, or WL):}
Baseline 0 with the weighted loss used in finetuning.

\textbf{Baseline 2 (resampling, or RS):}
Baseline 0 with the resampling technique used in finetuning.

\textbf{Ours:} Use the proposed two-stage training scheme, resampling and weighted loss are not applied.

\textbf{Ours (WL):} Use the proposed two-stage training scheme together with the weighted loss in the second stage.

\textbf{Ours (RS):} Use the proposed two-stage training scheme together with the resampling in the second stage.

By comparing our proposed method with Baselines 0, 1, and 2, we can verify if the proposed method can mitigate the data imbalance problem in the collected dataset and improve the overall performance of the classifier.

Due to the limited number of data samples in the dataset, we adopt the k-fold cross-validation~\cite{stone1974cross} technique to utilize all the data for more reliable performance estimation. In our experiment, we empirically set $k$ to 5 to balance the computation cost for each setting. See more details in the supplementary material.


 
\begin{figure}[b]
\centering
\includegraphics[width=0.45\textwidth]{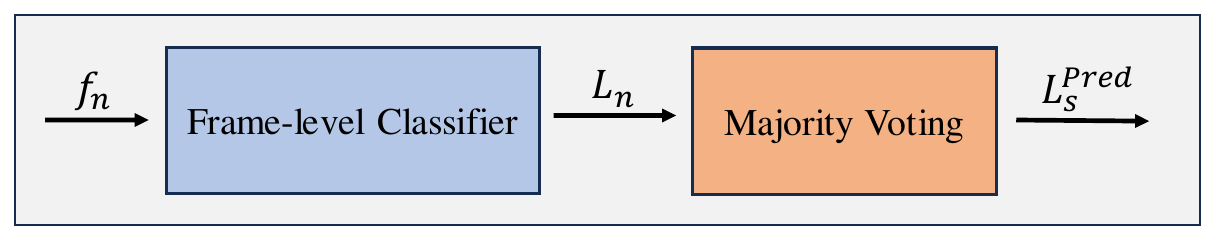}
\caption{Process of generating the class label for each meal sequence, $f_n\in\mathbb{R}^{F \times H \times W \times 3}$ is the frames within one meal, $L_n\in \mathbb{R}^{F \times 1}$ is the top-1 prediction for each frame and $L_s^{Pred}\in \mathbb{R}^{1}$ is the final label for this sequence}
\label{fig: Acc_seq}
\end{figure}
\subsection{Evaluation Metrics}
Since the goal of our method is to recognize the environment associated with eating activities, we use sequence-level accuracy instead of frame-level accuracy. The top-1 prediction result for each frame in a meal sequence is aggregated using majority voting to select the class label for this sequence (See Fig.~\ref{fig: Acc_seq}). 

The \textbf{sequence-level accuracy (SLA)} is defined as ~(\ref{eq:seq-accuracy})
\begin{equation}
\label{eq:seq-accuracy}
    \text{SLA} = \frac{S}{N}
\end{equation}
$S$ denotes the number of correctly classified sequences (meals), and $N$ denotes the total number of meals.

For the evaluation of the proposed ingestion environment recognition method, we use imbalance accuracy metrics~\cite{mortaz2020imbalance} such as macro-average precision, recall, and F1-score~\cite{mosley2013balanced}. See definitions in the supplementary material.

\subsection{Main Result}
\label{sec: main result}
\textbf{Overall performance:} The overall performance of the baseline methods and our proposed methods are summarized in Table~\ref{table:main result 1}. We report the sequence-level accuracy (after majority voting in each sequence) and the macro-average metrics for three different network backbones. The best results for each network backbone are bolded. 

As reported in Table~\ref{table:main result 1}, comparing the proposed two-stage training scheme (Ours) to Baselines 0, 1, and 2, it is noted that applying weight loss functions or resampling (Baseline 1 or Baseline 2) does not help improve the performance for normal finetuning (direct finetune on our dataset).  In contrast, our method achieves better overall performance with better sequence-level accuracy and much better macro-average metrics. As we see later in Table~\ref{table: main result minority}, Baselines 1 and 2 cannot predict the ``Vehicle" class correctly.

Due to severe data imbalance and limited training sample, simple re-weighting (WL/RS) can cause overfitting of limited ''Vehicle" class data for the feature classifier $g$.
However, by effectively incorporating more training samples from the Place365-ours dataset utilizing our proposed drop-then-maintain strategy, the severe data imbalance problem is alleviated with more training samples and the improvement of intra-class diversity in minority classes. Results (Ours) on all network backbones get a significant performance boost in all evaluation metrics. It is worth noting that the macro-average metrics of the model show even more improvement.

Ours (WL) and Ours (RS) achieve better performance in all metrics compared to Baselines 1, 2, and Ours, which shows a positive interaction between both approaches and our proposed framework. The aforementioned overfitting problem is alleviated here since more relevant training samples are included for training a less biased feature classifier using Place365-ours, which is made possible by the proposed semantic-based class filtering and merging. This result further verifies our previous argument of the necessity of the two-stage drop-then-maintain training strategy.

\textbf{Performance on the minority class:} Since our dataset has a severe data imbalance problem where the least represented class ``Vehicle" only contains 2 sequences, we want to further verify if our proposed method can help to improve the classification accuracy for it. We report the class accuracy for ``Vehicle" in Tabel~\ref{table: main result minority}. We can see that all of the Baseline methods 0, 1, and 2 fail to predict the minority class ``Vehicle". Even though baseline methods 1 and 2 have utilized weighted loss and resampling techniques to mitigate the negative influence of lacking training samples in the minority class, they still suffer from the overfitting problem since the number of training samples is too small for the model to capture the general characteristic of data distribution for ``Vehicle" class. 

However, by incorporating more relevant training samples from Place365-ours, the proposed method can classify the eating scene for the ``Vehicle" class. This is consistent through all network architectures. The large intra-class difference in our dataset is the main reason for the inability to correctly classify both two sequences labeled as ``Vehicle". In our dataset, one ``Vehicle" sequence is captured on the front seat of a family car while another sequence is captured on a bus. Since the second stage of training (on our dataset) uses one sequence for training and another for testing, the model learning to classify the environment on a family car may fail to classify the environment on a bus and vice versa.

The significant improvement in classification accuracy in the minority class can further verify that our proposed method helps handle the inherent data imbalance problem existing in eating scene recognition tasks.

\begin{table}[]
\caption{Main Result: Minority Class Accuracy}
\resizebox{0.475\textwidth}{!}{%
\begin{tabular}{c|cccccc}
\hline
                               & \multicolumn{6}{c}{Sequence-level Accuracy (\%) for Vehicle}                                                                              \\ \cline{2-7} 
                               & \multicolumn{3}{c|}{Baseline Method}                                               & \multicolumn{3}{c}{Our Method}                                  \\ \cline{2-7} 
\multirow{-3}{*}{Architecture} & \multicolumn{1}{c|}{DR} & \multicolumn{1}{c|}{WL} & \multicolumn{1}{c|}{RS} & \multicolumn{1}{c|}{Ours}  & \multicolumn{1}{c|}{+WL}   & +RS   \\ \hline
ResNet                         & \multicolumn{1}{c|}{0.00} & \multicolumn{1}{c|}{0.00}      & \multicolumn{1}{c|}{0.00}      & \multicolumn{1}{c|}{50.00} & \multicolumn{1}{c|}{50.00} & 50.00 \\ \hline
ConvNeXt                       & \multicolumn{1}{c|}{0.00} & \multicolumn{1}{c|}{0.00}      & \multicolumn{1}{c|}{0.00}      & \multicolumn{1}{c|}{50.00} & \multicolumn{1}{c|}{50.00} & 50.00 \\ \hline
Swin transformer                & \multicolumn{1}{c|}{0.00} & \multicolumn{1}{c|}{0.00}      & \multicolumn{1}{c|}{0.00}      & \multicolumn{1}{c|}{50.00} & \multicolumn{1}{c|}{50.00} & 50.00 \\ \hline
\end{tabular}%
}
\label{table: main result minority}
\end{table}

\begin{table}[]
\caption{Dataset Utilization of Training Strategies}
\centering
\resizebox{0.475\textwidth}{!}{%
\begin{tabular}{c|ccc}
\hline
\multirow{2}{*}{} & \multicolumn{3}{c}{Dataset Utilization}                                    \\ \cline{2-4} 
                  & \multicolumn{1}{c|}{\hspace*{1.5mm}ImageNet\hspace*{1.5mm}} & \multicolumn{1}{c|}{Place365-Ours} & \hspace*{2mm}Our dataset\hspace*{2mm} \\ \hline
Strategy 1        & \multicolumn{1}{c|}{}         & \multicolumn{1}{c|}{}              & \checkmark      \\ \hline
Strategy 2        & \multicolumn{1}{c|}{\checkmark}         & \multicolumn{1}{c|}{}              & \checkmark     \\ \hline
Strategy 3        & \multicolumn{1}{c|}{\checkmark}         & \multicolumn{1}{c|}{\checkmark}              &       \\ \hline
Strategy 4        & \multicolumn{1}{c|}{\checkmark}         & \multicolumn{1}{c|}{\checkmark}              & \checkmark      \\ \hline
\end{tabular}%
}
\label{table: strategy difference}
\end{table}

\begin{table}[]
\caption{Different Training Strategies: Overall Performance}
\centering
\resizebox{0.475\textwidth}{!}{%
\begin{tabular}{c|cc|cc|cc}
\hline
\multirow{2}{*}{} & \multicolumn{2}{c|}{ResNet}        & \multicolumn{2}{c|}{ConvNeXt}      & \multicolumn{2}{c}{Swin transformer} \\ \cline{2-7} 
                  & \multicolumn{1}{c|}{Acc}   & F1    & \multicolumn{1}{c|}{Acc}   & F1    & \multicolumn{1}{c|}{Acc}    & F1    \\ \hline
Strategy   1      & \multicolumn{1}{c|}{77.53} & 39.34 & \multicolumn{1}{c|}{73.03} & 28.74 & \multicolumn{1}{c|}{78.65}  & 42.28 \\ \hline
Strategy   2      & \multicolumn{1}{c|}{85.39} & 56.87 & \multicolumn{1}{c|}{88.76} & 61.15 & \multicolumn{1}{c|}{88.76}  & 77.21 \\ \hline
Strategy   3      & \multicolumn{1}{c|}{82.02} & 69.42 & \multicolumn{1}{c|}{89.89} & 79.20 & \multicolumn{1}{c|}{89.89}  & 81.02 \\ \hline
Strategy   4      & \multicolumn{1}{c|}{\textbf{89.89}} & \textbf{78.00} & \multicolumn{1}{c|}{\textbf{92.13}} & \textbf{80.17} & \multicolumn{1}{c|}{\textbf{94.38}}  & \textbf{85.34} \\ \hline
\end{tabular}%
}
\begin{tablenotes}
    \footnotesize
    \item F1 score is the macro-average F1 score calculated from the confusion matrix for each result. 
\end{tablenotes}
\label{table:ablation studies overall}
\end{table}

\begin{table}[]
\caption{Maintaining Strategy for Feature Classifier}
\centering
\resizebox{0.475\textwidth}{!}{%
\begin{tabular}{l|cc|cc|cc}
\hline
\multirow{2}{*}{} & \multicolumn{2}{c|}{ResNet}        & \multicolumn{2}{c|}{ConvNeXt}      & \multicolumn{2}{c}{Swin transformer} \\ \cline{2-7} 
                  & \multicolumn{1}{c|}{Acc}   & F1    & \multicolumn{1}{c|}{Acc}   & F1    & \multicolumn{1}{c|}{Acc}    & F1    \\ \hline
Ours w/o M       & \multicolumn{1}{c|}{88.76} & 61.67 & \multicolumn{1}{c|}{85.39} & 54.69 & \multicolumn{1}{c|}{87.64}  & 58.96 \\ \hline
Ours w/ M      & \multicolumn{1}{c|}{89.89} & 78.00 & \multicolumn{1}{c|}{92.13} & 80.17 & \multicolumn{1}{c|}{94.38}  & 85.34 \\ \hline
\end{tabular}%
}
\begin{tablenotes}
    \footnotesize
    \item F1 score is the macro-average F1 score calculated from the confusion matrix for each result. M denotes the maintaining of the feature classifier.
\end{tablenotes}
\label{table:ablation studies labelmerge}
\end{table}


\subsection{Ablation Studies}
\label{sec: ablation studies}
\textbf{Additional datasets:} 
Since we have two additional datasets (ImageNet and Place365-Ours) that are used for training our model, there are four training strategies presented in the experiments to explore the effectiveness of including the additional datasets.
Table~\ref{table: strategy difference} summarizes the differences between the four training strategies. See the description of strategies in the supplementary material.





From Table~\ref{table:ablation studies overall}, we observe that strategy 4, the proposed method, is the best training strategy since it achieves the highest accuracy and Macro-Average F1 score for all network architectures. This is expected since the proposed two-stage training strategy optimally utilizes two additional datasets as well as our dataset for improving performance. 

\textbf{Maintaining of feature classifier:} 
As mentioned in Section~\ref{sec: model formulation}, we maintain the feature classifier $g$ from stage 1 to stage 2 instead of dropping it (enabled by semantic-based class filtering and merging). We verify the strategy and see results in Table~\ref{table:ablation studies labelmerge}.

We observe a significant improvement in sequence-level accuracy and even more improvement in macro-average F1 score by maintaining the feature classifier $g$ for the second stage of training. The classifier is much more robust to data imbalance issues, demonstrating the effectiveness of our proposed semantic-based label filtering and merging that enables the drop-then-maintain strategy.
\section{Conclusion}
\label{sec:conclusion}
In this paper, we present an automatic ingestion environment recognition method to aid nutritionists and dietitians in overcoming the limitations of self-report and manual review of eating scene images. We use the data from the accelerometer and flexible sensor to detect eating episodes and explore the proposed two-stage drop-then-maintain training framework on several neural network architectures to perform scene classification. The experimental results indicate our proposed method outperforms the baseline methods in both sequence-level accuracy and minority-class accuracy, demonstrating its effectiveness in addressing the data imbalance problem.

A larger study is planned for the future to extend the work to more environment categories for a longer duration of device wear. At the same time, a major advantage of this study is that the image database is fully natural and contains no staged images or artificially created environments. Therefore, the performance of the proposed method is representative of what is expected in everyday living.

\clearpage
{
    \small
    \bibliographystyle{ieeenat_fullname}
    \bibliography{main}
}


\end{document}